\documentclass{iaus}
\usepackage{graphicx}
\usepackage{array}

\title[Spitzer Observations of PNe] 
{Spitzer Observations of Planetary Nebulae}

\author[Y.-H.\ Chu]   
{You-Hua Chu}

\affiliation{Department of Astronomy, University of Illinois, \\ 1002 West Green Street,
Urbana, Illinois, 61801, USA \\ email: {\tt yhchu@illinois.edu}}

\pubyear{2011}
\volume{283}  
\pagerange{1-8}
\jname{Planetary Nebulae: An Eye to the Future}
\editors{A.\ Manchado, L.\ Stanghellini, \& D.\ Sch\"onberner, eds.}
\begin{document}

\maketitle

\begin{abstract}
The Spitzer Space Telescope has three science instruments
(IRAC, MIPS, and IRS) that can take images at 
3.6, 4.5, 5.8, 8.0, 24, 70, and 160 $\mu$m, spectra over
5--38 $\mu$m, and spectral energy distribution over 52--100 $\mu$m.
The Spitzer archive contains targeted imaging observations for more
than 100 PNe.  Spitzer legacy surveys, particularly the GLIMPSE 
survey of the Galactic plane, contain additional serendipitous imaging 
observations of PNe.  Spitzer imaging and spectroscopic observations 
of PNe allow us to investigate atomic/molecular line emission and 
dust continuum from the nebulae as well as circumstellar dust disks 
around the central stars.  Highlights of Spitzer observations of 
PNe are reviewed in this paper.
\end{abstract}

\firstsection 
\section{Spitzer Space Telescope}

The Spitzer Space Telescope (Werner et al.\ 2004), one of the four NASA's 
Great Observatories, is a 0.85 meter diameter, f/12 infrared (IR) 
telescope launched on August 25, 2003.  It has an Earth-trailing 
heliocentric orbit so that the telescope is kept away from the Earth's heat
and can be cooled more efficiently. 

Spitzer has three science instruments: \\
(1) The Infrared Array Camera (IRAC; Fazio et al.\ 2004) has four detectors 
that take images at 3.6, 4.5, 5.8, and 8.0 $\mu$m, respectively. \\
(2) The Multiband Imaging Photometer for Spitzer (MIPS; Rieke et al.\ 2004)
has three detector arrays for imaging at 24, 70, and 160 $\mu$m, respectively,
and the 70 $\mu$m detector can also be used to measure spectral energy 
distributions (SEDs) from 52 to 100 $\mu$m at a spectral resolution of $\sim$7\%. \\
(3) The InfraRed Spectrograph (IRS; Houck et al.\ 2004) has four modules 
that allow us to take spectra with a high  
($\lambda/\Delta\lambda \sim$600) or low ($\lambda/\Delta\lambda \sim$64--128)
resolution over the spectral range of 5--38 $\mu$m. \\
When the liquid helium coolant was exhausted, the ``Warm Spitzer'' phase started 
in May 2009, and only IRAC detectors at 3.6 and 4.5 $\mu$m remained useful.

Spitzer observations of PNe have been obtained by Guaranteed Time Observers (GTOs),
Guest Observers (GOs), or legacy surveys of the Galactic plane and the 
Large and Small Magellanic Clouds (LMC \& SMC).  
The GTOs are associated with the Spitzer instrument builders, and the 
GOs are open to the public.  The surveys of the Galactic plane have been made 
for $l = \pm$65$^\circ$ and $b = \pm$1$^\circ$ in the four IRAC bands (GLIMPSE I
and GLIMPSE II) and in MIPS 24 $\mu$m (MIPSGAL I and MIPSGAL II).  The IRAC
observations were extended to higher galactic latitudes for selected
longitudes (GLIMPSE 3D) in all four bands, and extended to cover the rest of
the Galactic plane in 3.6 and 4.5 $\mu$m during the Warm Spitzer mission 
(GLIMPSE360).  A large number of PNe are expected to be covered by these
GLIMPSE and MIPSGAL observations.
The LMC and SMC were surveyed in all IRAC and MIPS bands (SAGE-LMC; SAGE-SMC;
S$^3$MC).  All PNe, except the outlying members, in the LMC and SMC were
included in these surveys.

\section{IRAC Observations of PNe}

The four IRAC bands contain different combinations of polycyclic aromatic
hydrocarbon (PAH) features, HI recombination lines, H$_2$ rotational 
transitions, and forbidden lines of atoms/ions, as marked in Table 1 and
seen in the ISO spectra of NGC 7027 (Bernard-Salas et al.\ 2001) and Hb 12
(Hora et al.\ 2004).  To interpret IRAC images, spectra or images at other 
wavelengths are needed.  Nevertheless, IRAC images often reveal morphological
features that bear implications on PN formation mechanisms but are less well 
seen in other wavelengths.  These features include:
(1) Molecular knots -- the Helix Nebula is the best example and the 
molecular nature of its knots has been confirmed spectroscopically
by Hora et al.\ (2006); IRAC 8 $\mu$m images of A21 and JnEr 1 in 
Figure 1 also show molecular knots. (2) Large halos -- IRAC 8 $\mu$m
images frequently show prominent, extended halos similar to those
seen in optical emission lines (Corradi et al.\ 2003); six new halos are
shown in Figure 2.  (3) Concentric rings -- IRAC images of PNe also reveal
concentric rings similar to those seen in optical emission lines 
(Corradi et al.\ 2004); two examples are shown in Figure 3.
(4) Circular rings in halos -- IRAC 8 $\mu$m images of NGC 6072 and NGC 6720
in Figure 3 show an almost perfectly circular ring in the halo that is not 
obvious in optical images.  (5) PAH emission peaks of ionization bounded
nebulae -- PAH emission can peak in the photodissociation regions (PDRs)
around the photoionized PN, for example, NGC 2610 in Figure 4. 

\begin{table}[t]
  \begin{center}
  \caption{Spectral Features in the IRAC Bands}
  \label{tab1}
\begin{tabular}{lcccc}
\hline
IRAC       &  PAH         &   HI   &   H$_2$ & Ion \\
Band       &              & Recom. &  Rot.   & Forb.\\
\hline
3.6 $\mu$m &  $\checkmark$& $\checkmark$  &   &  \\
4.5 $\mu$m &              & $\checkmark$& $\checkmark$  &  $\checkmark$    \\
5.8 $\mu$m &  $\checkmark$& $\checkmark$& $\checkmark$  &  $\checkmark$    \\
8.0 $\mu$m &  $\checkmark$& $\checkmark$& $\checkmark$  &  $\checkmark$    \\
\hline
 \end{tabular}
  \end{center}
\end{table}

\begin{figure}[b]
\begin{center}
\includegraphics[width=4.5in]{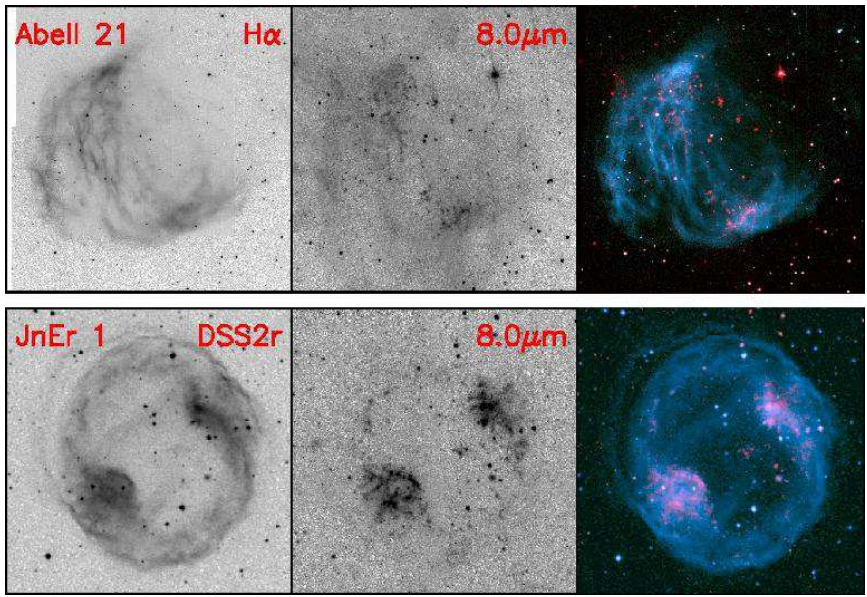}
\caption{Images of PNe A21 and JnEr 1 illustrating molecular knots.}
\label{fig1}
\end{center}
\end{figure}

\begin{figure}
\begin{center}
\includegraphics[width=4.5in]{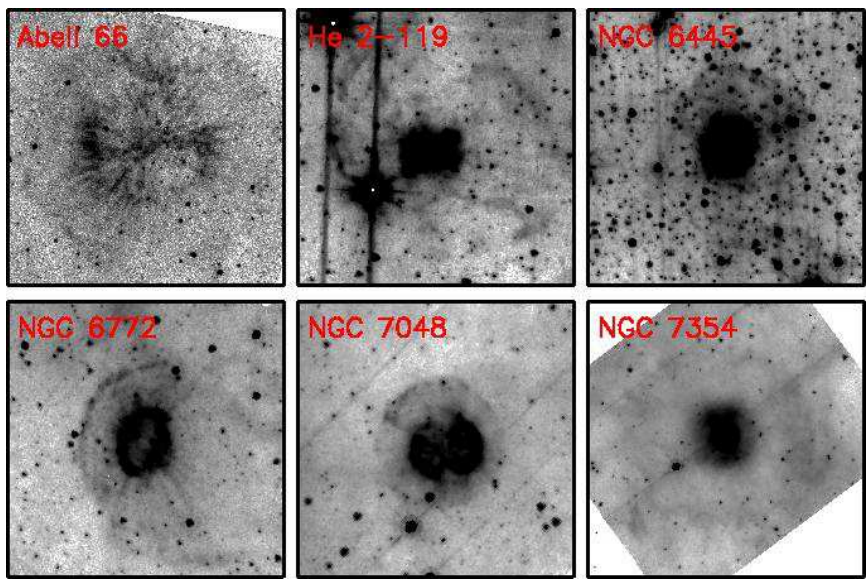}
\caption{IRAC 8 $\mu$m images of PNe A66, He 2-119, NGC 6445, NGC 6772, 
NGC 7048, and NGC 7354 showing large halos.}
\label{fig2}
\end{center}
\end{figure}

\begin{figure}
\begin{center}
\includegraphics[width=5in]{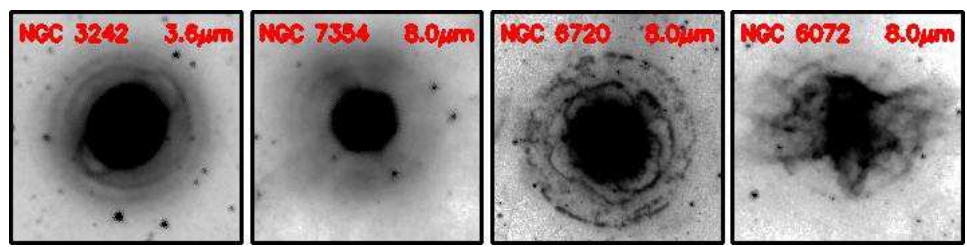}
\caption{IRAC 3.6 $\mu$m image of NGC 3242 and 8.0 $\mu$m image of NGC 7354
showing concentric rings, and IRAC 8.0 $\mu$m images of NGC 6720 and NGC 6072
showing an almost perfectly circular ring in the halo.}
\label{fig3}
\end{center}
\end{figure}

\begin{figure}
\begin{center}
\includegraphics[width=2.5in]{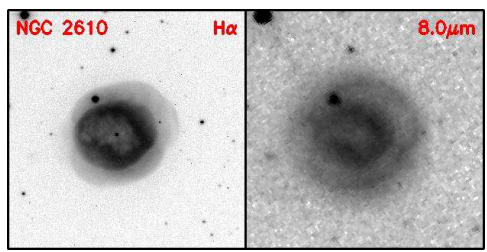}
\caption{H$\alpha$ and IRAC 8.0 $\mu$m images of NGC 2610 showing the
PAH emission along the outer rim.}
\label{fig4}
\end{center}
\end{figure}

Since the initial report of IRAC observations of PNe (Hora et al.\ 2004),
not many analyses of spatially-resolved IRAC images of PNe have been published.
A detailed analysis of IRAC images of the Helix Nebula complemented
by spectroscopic observations was carried out by Hora et al.\ (2006).
Phillips \& Ramos-Larios (2008a) analyzed quantitatively IRAC images
of 18 PNe and suggest that PAH emission from PDRs 
contributes significantly to the observed mid-IR fluxes.
Phillips \& Ramos-Larios (2010) extended their analysis to seven bipolar
PNe, assessing the roles of PAH emission from PDRs and shock excited 
H$_2$ emission, and discussing the color variations from the cores to
the lobes.  Cerrigone et al.\ (2008) analyzed IRAC images, in conjunction
with 4.8 and 8.6 GHz radio continuum observations, of the bipolar PN
IC 4406, and concluded that there exist at least three dust components at 
temperatures ranging from 57 to 700 K.

The GLIMPSE surveys have been used to search for new PNe.  Phillips \&
Ramos-Larios (2008b) found 12 new and possible PNe in GLIMPSE I.
Kwok et al.\ (2008) also used GLIMPSE I survey data to search for
new PNe; however, about 1/3 of the 30 PNe they suggested are likely
HII regions in the Galactic plane.  Zhang \& Kwok (2009) used GLIMPSE II,
complemented by MIPS 24 $\mu$m and H$\alpha$ images, to identify 37 
new PNe.  Finally, Quino-Mendoza et al.\ (2011) have used the GLIMPSE3D 
survey to make quantitative analyses of the color and surface brightness 
variations in 24 PNe.

IRAC observations of PNe have been used to investigate how their
IR colors can be used to identify new PNe.  Using different combinations
of IRAC bands, such as [3.6]$-$[4.5], [4.5]$-$[8.0], and [5.8]$-$[8.0],
color-color diagrams have been constructed for 12 Galactic PNe
(Hora et al.\ 2004), 233 PNe in the LMC (Hora et al.\ 2008), 58 PNe from
the MASH sample (Cohen et al.\ 2007), 136 PNe in the GLIMPSE I survey
(Cohen et al.\ 2011), and 367 PNe from the Galaxy and the LMC
(Phillips \& Ramos-Larios 2009).  As shown in the left panel of Figure 5
and concluded by Hora et al.\ (2008) and Gruendl \& Chu (2009), PNe are 
well separated from stars in color-color and color-magnitude diagrams, 
but overlap significantly with young stellar objects and background galaxies.  
To unambiguously identify PNe, morphological information of the nebulae 
and their environments and/or spectra are still needed.  An interesting 
color evolution, shown in the right panels of Figure 5, has been noted by 
Cohen et al.\ (2011); the IRAC 5.8 $\mu$m band flux increases with PN size
and it is attributed to an increasing contribution from the 6.2 $\mu$m PAH 
band emission as PNe age.

\begin{figure}
\begin{center}
\includegraphics[width=3in]{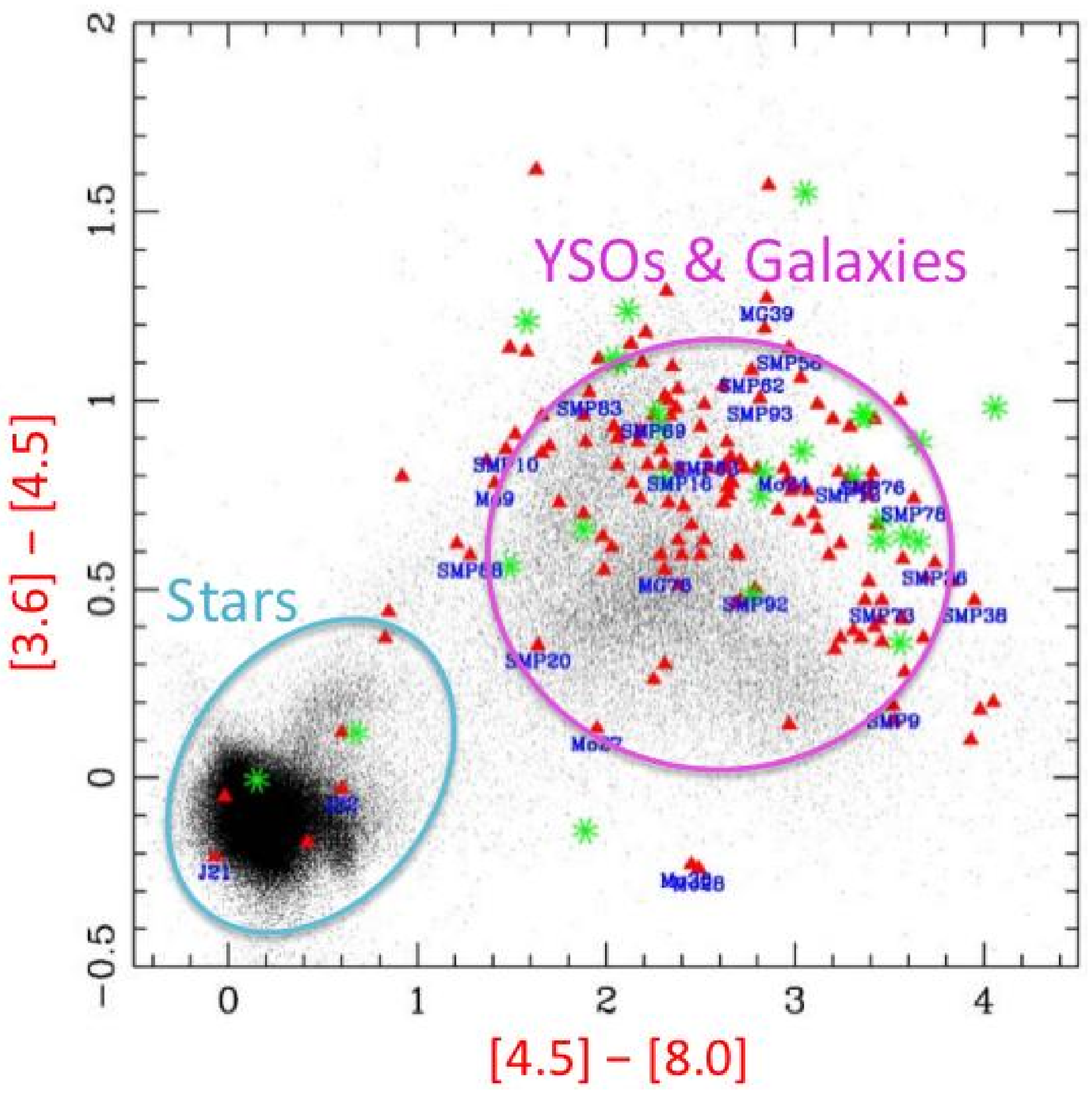}~~
\includegraphics[width=1.85in]{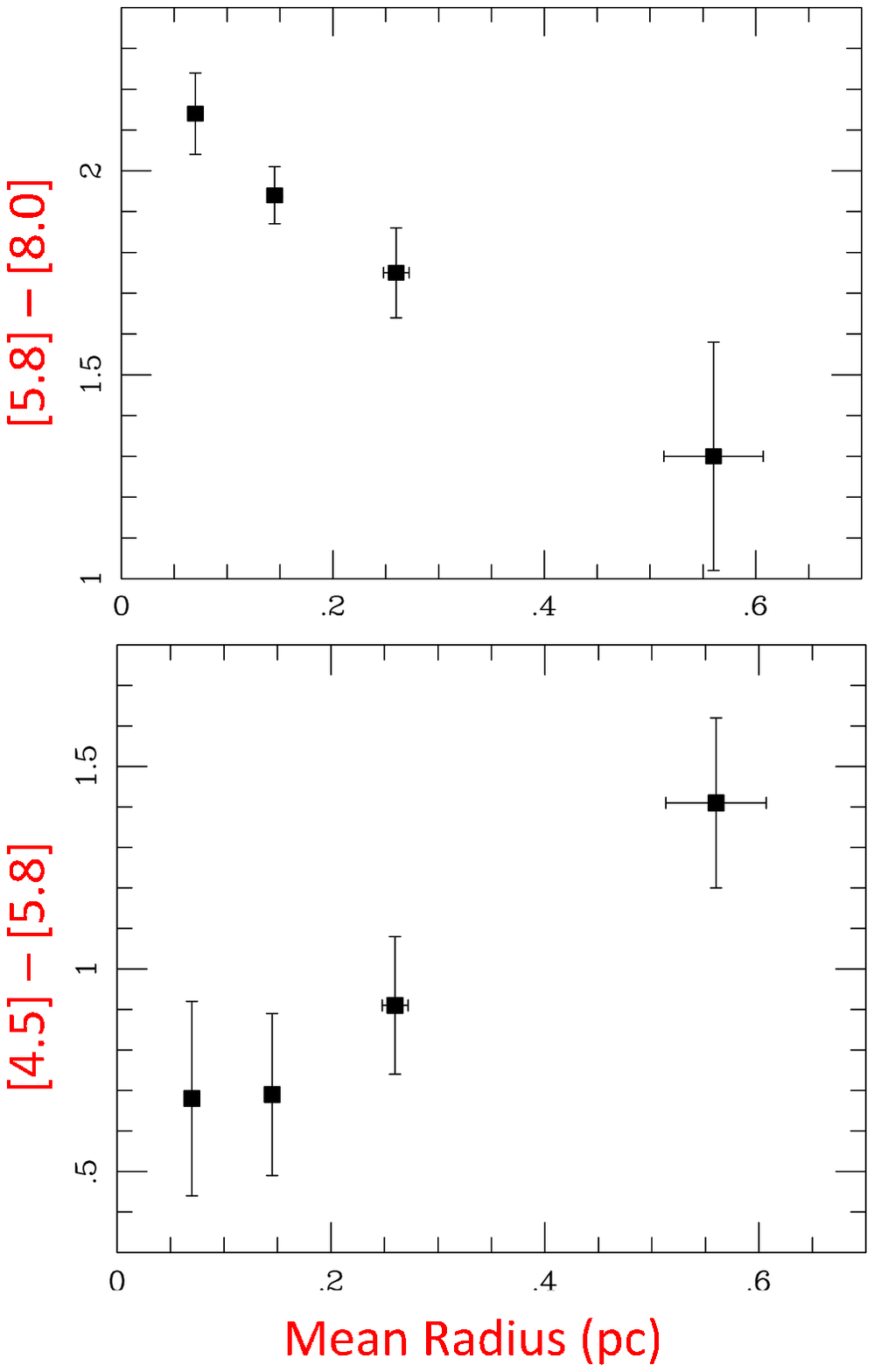}
\caption{{\bf Left}:Color-color diagram of sources in the LMC (Hora et al.\ 2008).  
The triangles mark known PNe in the LMC, and asterisks Galactic PNe. 
The ellipse at the lower left encloses locations of stars, and the upper 
ellipse encloses the general locations of young stellar objects (YSOs) and 
background galaxies. {\bf Right}: Color-size plots of PNe from GLIMPSE I
(Cohen et al.\ 2011).}
\label{fig5}
\end{center}
\end{figure}

\section{MIPS Observations of PNe}

The three MIPS bands, 24, 70, and 160 $\mu$m, are ideal for 
investigation of dust continuum emission; however, their 
point-spread functions, FWHM 6$''$, 18$''$, and 40$''$, respectively,
severely limit the number of PNe that are adequately resolved 
in the 70 and 160 $\mu$m bands for detailed analysis.
Spatially resolved analyses of PNe utilizing all MIPS bands
have been reported for only two bipolar nebulae, NGC 2346 and NGC 650.
For NGC 2346, Su et al.\ (2004) show that the 24 $\mu$m image exhibits
a bipolar morphology resembling that seen in the optical [N II] image, 
the 70 $\mu$m emission is concentrated in a torus where the bipolar lobes
join, and the 160 $\mu$m emission is much more extended and showing
spherical symmetry.  They suggest that the mass loss changed from 
spherical symmetric to non-spherical due to the binary central star
of NGC 2346.  For NGC 650, Ueta (2006) finds the 24 $\mu$m image 
similar to the [O III] image and suggests that the [O~IV] 25.89 $\mu$m
line emission dominates the MIPS 24 $\mu$m band flux; he finds the
70 and 160 $\mu$m emission in the torus/waist more extended than 
the 24 $\mu$m emission and suggests that mass loss was mainly in the
equatorial, rather than polar, directions.

MIPS 24 $\mu$m images of PNe are useful in revealing regions of
high ionization as well as dust emission. 
The MIPS 24 $\mu$m band includes high-ionization [O IV] 25.89 $\mu$m
and [Ne~V] 24.3 $\mu$m lines and dust continuum that are frequently seen
in PNe. Large numbers of MIPS 24 $\mu$m observations of PNe have been
reported by Phillips \& Marguez-Lugo (2011) and Chu et al.\ (2009).
The former investigation used the MIPSGAL data (each $\sim$30~s exposure) to 
analyze 224 PNe in the Galactic plane, comparing the 24 $\mu$m profiles 
to IRAC profiles and constructing color-color diagrams.  The latter 
obtained 24 $\mu$m images for 36 PNe each with 420~s exposure, 
compared the 24 $\mu$m profiles to H$\alpha$ profiles, and obtained IRS
spectra to probe the nature of the 24 $\mu$m emission.


H$\alpha$ emission is an excellent tracer for dense ionized gas.
Comparing 24 $\mu$m and H$\alpha$ profiles, Chu et al.\ (2009) 
find three types of 
correspondences, as illustrated in Figure~6: (1) 24 $\mu$m emission
more extended than H$\alpha$ emission -- as the [O IV] and [Ne V]
lines originate from high-ionization regions, the 24 $\mu$m emission
from the outer parts of PNe must be dominated by dust continuum;
(2) similar morphology and spatial extent between 24 $\mu$m and H$\alpha$
-- both line emission and dust continuum can contribute to the 24
$\mu$m band flux;  (3) 24 $\mu$m emission at center of H$\alpha$ shell 
-- low densities and high excitation are expected in central regions of PNe,
and thus the 24 $\mu$m emission must be dominated by [O IV] and some 
[Ne V] line emission.  IRS spectrum of the diffuse  24 $\mu$m emission
in the central region of Sh 2-188 shows that the emission is indeed
dominated by the [O IV] 25.89 $\mu$m emission (Chu et al.\ 2009).

\begin{figure}[h]
\begin{center}
\includegraphics[width=4.5in]{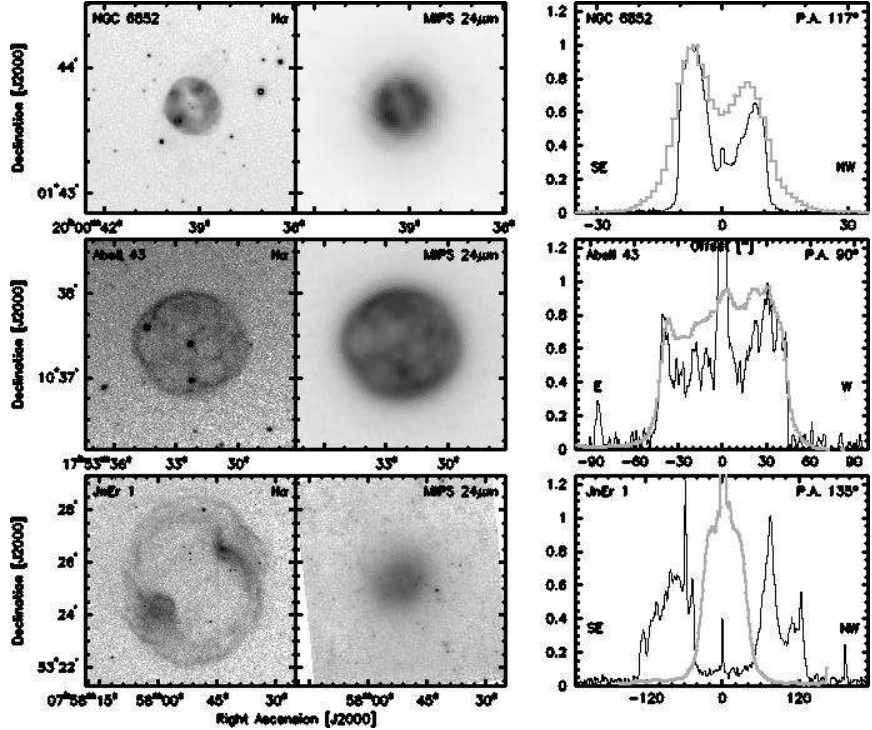}
\caption{H$\alpha$ and 24 $\mu$m images of NGC 6852, A43, and JnEr 1.
The right panels compare the H$\alpha$ (dark) and 24 $\mu$m (grey) 
surface brightness profiles.  Taken from Chu et al.\ (2009).}
\label{fig6}
\end{center}
\end{figure}

\section{IRS Observations of PNe}

A picture is worth a thousand words, while a spectrum
is worth a thousand pictures.  The most exciting Spitzer
results of PNe are derived from IRS observations,
which have many advantages over spectroscopic observations
in optical wavelengths for studies of abundances of
ionized gas, exciting/fundamental molecules, and dust properties.

IRS studies of abundances of ionized gas in PNe have the
following advantages: (1) extinction has little effects at
mid-IR wavelengths; (2) mid-IR transitions are close to the
ground level and thus insensitive to the temperature;
(3) most ionization stages of Ne, S, and Ar are covered
and thus there is no need to resort to ionization correction
factors; and (4) [Fe III] 22.93 $\mu$m, [Fe II] 17.94 $\mu$m,
[Cl IV] 20.32 and 11.76 $\mu$m, and [P~III] 17.89~$\mu$m lines
allow determinations of Fe, Cl, and P abundances.  

Therefore, many Galactic PNe have been
observed for abundance analyses, and the nebular abundances
are further compared to models of stellar evolution and
nucleosynthesis to determine the mass of progenitor stars:
IC 2448, M 1-42, NGC 2392, NGC 2792, NGC 3242, 
NGC 6210, NGC 6369 (Guiles et al.\ 2007; Pottasch et al.\ 2008, 
2009a, 2009b; Pottasch \& Bernard-Salas 2008).  
Abundances of the halo PN DdDm-1 have been analyzed by
Henry et al.\ (2008) and the most O-deficient PN TS01 by
Stasi\'nska et al.\ (2010).  See Kwitter (this volume) for
a review of PN abundances.

Bernard-Salas et al.\ (2004) analyzed the abundances of 
the LMC PN SMP 83.  Bernard-Salas et al.\ (2008) analyzed
Ne and S abundances for 25 PNe in the LMC and SMC, finding 
LMC and SMC Ne abundances to be $\sim$1/3 and 1/6 of the 
Galactic values, respectively; furthermore, the Ne/S ratio 
is higher in the MC PNe, $\sim$23.5, than in the Galactic PNe,
$\sim$16.  Shaw et al.\ (2010) analyzed elemental abundances of
He, N, O, Ne, S, and Ar for 14 PNe in the SMC, and found that
ionization correction factors work well even in low-metalicity
conditions, except for PNe with very high ionization.

The most exciting discovery in PNe made by IRS observations 
is the detection of fullerenes. C$_{60}$ and C$_{70}$ were produced 
in laboratory more than 25 years ago (Kroto et al.\ 1985), and
they might be responsible for the diffuse interstellar bands
and provide catalyst for prebiotic chemistry.  However,
they have not been unambiguously detected until 
IRS observations of PNe were made (Figure 7).  
Cami et al.\ (2010) reported 
the first detection of C$_{60}$ and C$_{70}$ in the PN TC 1, and
suggested that fullerenes can be produced only in H-deficient 
environments.  However, this suggestion has been challenged by
Garc\'ia-Hern\'andez et al.\ (2010), who detected fullerene in 
H-containing PNe in the Galaxy and in the SMC.  The simultaneous
presence of PAHs and fullerenes is suggestive that they may both
be formed by photochemical processing of hydrogenated 
amorphous carbons.

\begin{figure}
\begin{center}
\includegraphics[width=4.1in]{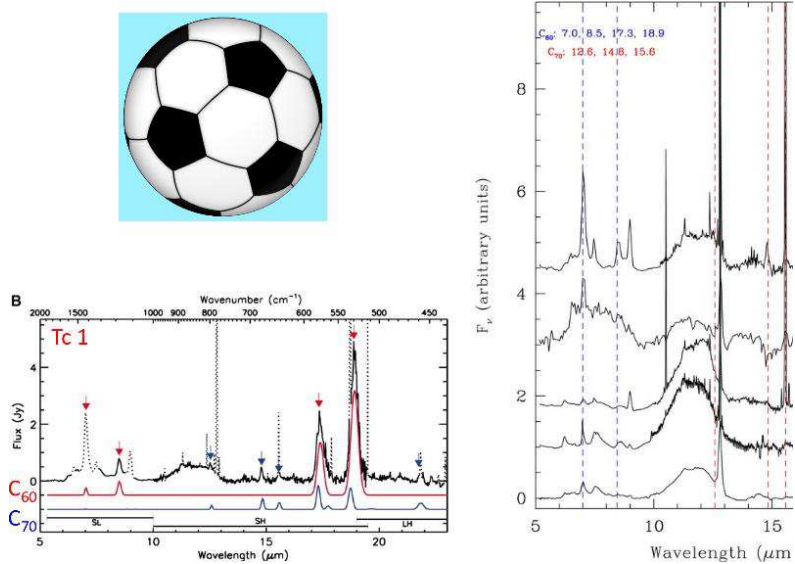}
\caption{Spectra of fullerene in PNe.  The left panel is from
Cami et al.\ (2010) and the right panel from Garc\'ia-Hern\'andez 
et al.\ (2010).}
\label{fig7}
\end{center}
\end{figure}

IRS spectra of 11 PNe in the Galactic bulge have been studied
by Gutenkunst et al.\ (2008) -- these spectra are dominated by 
O-rich dust features (crystalline silicates) that are rare in 
disk PNe; furthermore, roughly 1/2 of these bulge PNe also show
C-rich dust features (PAHs), i.e., dual chemistry.  They suggest
that the dual chemistry is caused by binary evolution, with the
PAHs located in the current outflow and the crystalline silicates 
in an old disk created by binary interaction.  This mixed chemistry
phenomenon in Galactic bulge PNe was re-examined by Perea-Calder\'on 
et al.\ (2009), who analyzed 40 Galactic PNe of which 26 belong to 
the bulge.  These authors propose a different explanation for the
dual chemistry -- the final thermal pulse of the AGB produced enhanced
mass loss, removing the remaining H-rich envelope, exposing the 
C-rich layer, and generating shocks to produce the crystalline 
silicates.

Stanghellini et al.\ (2007; this volume) made IRS observations
of a large number of PNe in the LMC and the SMC.  These observations
will be complemented by images and long-slit spectra obtained with
the Hubble Space Telescope as well as ground-based telescopes.  
As the distances to these PNe are well known, it is possible to
analyze dust properties of these PNe and compare them with the
evolutionary status of their central stars for the most comprehensive
picture.

\section{Spitzer Observations of CSPNs}

Spitzer MIPS observations of the Helix Nebula show a bright
unresolved source coincident with the central star in both
the 24 and 70 $\mu$m bands.  Followup IRS observations confirm
the dust continuum nature of this excess mid-IR emission.
The color temperature of the IR emitter is 90-130 K, too
cold to be a star.  The IR luminosity requires an emitting 
area of 4-40 AU$^2$, which can be provided only by a dust
disk.  For a stellar temperature of $\sim$110,000 K,
the dust disk must be 35-150 AU from the central star.
As such a distance corresponds to the Kuiper Belt in the
Solar System, Su et al.\ (2007) suggest that the dusk disk
is produced by collisions among Kuiper Belt-like objects.

To search for dust disks similar to that of the Helix
central star, Chu et al.\ (2011) made a MIPS 24 $\mu$m 
survey of 71 hot white dwarfs, about 1/2 of which are 
still in PNe.  They find 24 $\mu$m excesses in 9 objects,
among which 7 are in PNe.  Inspired by the prevalence of 
IR excesses in central stars of PNe, Bilikova et al.\
(this volume) made an Spitzer archival search for IR 
excesses of PN central stars and made followup spectroscopic
observations using Spitzer IRS and Gemini NIRI and
Michelle.  While spectroscopic observations show
that most of the IR excesses are dominated by dust 
continuum, some also show emission lines or even
silicate features.  These dust disks may have resulted 
from planetary disk evolution or binary central
star interactions.

\end{document}